\documentstyle[11pt,aaspp4]{article}

\def\et{et al.}
\def\kms{km s$^{-1}$}

\def\solar{\ifmmode_{\mathord\odot}\;\else$_{\mathord\odot}\;$\fi}

\begin{document}

\title{Identifying Old Tidal Dwarf Irregulars}

\author{Deidre A.\ Hunter, Sally D.\ Hunsberger, and Erin W.\ Roye\footnote{
\rm Current address Yale University, New Haven, Connecticut 06520 USA}
}
\affil{Lowell Observatory, 1400 West Mars Hill Road, Flagstaff, Arizona 86001
USA;
\\dah@lowell.edu, sdh@lowell.edu, erin.roye@yale.edu}

\begin{abstract}

We examine the observational consequences of the two possible origins for
irregular galaxies: formation from collapse of a primordial cloud of gas early 
in the age of the Universe, and formation from tidal tails in an interaction
that could have occured any time in the history of the Universe.
Because the formation from tidal tails could have occurred a long time ago,
proximity to larger galaxies is not sufficient to distinguish tidal dwarfs
from traditional dwarfs.
We consider the effects of little or no dark matter on rotation speeds
and the Tully-Fisher
relationship, the metallicity-luminosity relationship, structure, and
stellar populations. From these selection criteria, we identify a small
list of dwarf irregular galaxies that are
candidates for having formed as tidal dwarfs.

\end{abstract}


\section{Introduction}

Models of interacting galaxies have shown that tidal forces in 
the interaction
can produce long tails of stars and gas that have been pulled out of the
interacting galaxies (Toomre \& Toomre 1972).
Zwicky (1956) pointed out the possibility that self-gravitating objects might
develop in these tidal tails that could then evolve into small galaxies.
Since then, concentrations of stars and gas that are probable ``tidal dwarfs''
in the making have been observed at the tips
of tidal tails in several interacting systems (for example, the
Antennae system; Mirabel \et\ 1992).
Numerical modelling
confirms that bound, galaxy-sized clumps can form in tidal tails
(Barnes \& Hernquist 1992;
Elmegreen, Kaufman, \& Thomasson 1993).

These tidal dwarfs, once the tidal tail itself has dispersed and the parent
galaxies have moved off, could bear
a striking resemblance to small, independent, Im-type galaxies
(Schweizer 1974).
The tidal dwarfs are small, gas-rich, morphologically disorganized,
and already have on-going star formation
(see also Mirabel, Lutz, \& Maza 1991).
Furthermore, the properties measured for tidal dwarfs
are well within the range of properties seen for
normal, relatively isolated irregular galaxies 
(Mirabel, Dottori, \& Lutz 1992; Duc \& Mirabel 1994; Hibbard
\et\ 1994; Hunter 1997).

Gravitational interactions are an on-going process in the Universe 
that began when galaxies
themselves first formed. Therefore, this mechanism for forming 
irregular galaxies
has been taking place for the age of the Universe.
Hunsberger, Charlton, \& Zaritsky (1996), for example, estimate that
as many as one-half of the current dwarf galaxy population of compact groups
may have been formed from the interactions of giant spiral galaxies.
The formation of dwarf irregulars in compact groups is accelerated
because of the increased crowdedness and potential for interactions there.
However, interactions can and do occur outside of compact groups
of galaxies
as well.

One must then consider that any given dwarf irregular 
galaxy, including field galaxies, could have been
formed in one of two ways: traditional formation from collapse of
a primordial
cloud of gas early in the age of the Universe, and tidal dwarf formation from
an interaction of larger galaxies at any time during
the history of the Universe. Because the time scale since the formation
of a tidal dwarf can be large, a tidal dwarf 
could appear to be relatively isolated if the formation
took place many Gyrs ago. 

Because the formation mechanism of traditional dwarfs and of tidal dwarfs
are different, some key characteristics of these two groups of galaxies
could also be different, as outlined by Barnes \& Hernquist (1992)
and Elmegreen \et\ (1993).
In this paper we examine a sample of irregular galaxies with these 
observational differences in mind and ask whether any seemingly normal
Im galaxy might be a candidate for an old tidal dwarf. The observational
characteristics of Im galaxies and distinctions with tidal dwarfs
are too imprecise at this time to do more than point out candidate
tidal dwarf systems, but it is a way to begin thinking about the issue
that not all irregulars may have had the same initial conditions.

Dwarf galaxies, galaxies that are intrinsically not luminous,
come in a variety of types, including spirals, irregulars, ellipticals,
and spheroidals,
and Gerola, Carnevali, \& Salpeter (1983)
suggested galaxy interactions as a means for forming
dwarf ellipticals. However,
the most common type of dwarf is a gas-rich Im galaxy, and we will primarily
concentrate on irregulars in this discussion.

\section{Selection Criteria for Old Tidal Dwarfs}

Since tidal dwarfs form from material drawn out of larger
galaxies, their properties could differ significantly
from those of traditional dwarf galaxies. Barnes \& Hernquist (1992)
suggest that tidal dwarfs formed from parent galaxies with dark matter
distributed in 
extended massive haloes (Barnes 1992) 
will have $<$5\% of the dwarf's mass in dark matter.
Thus, tidal dwarfs should have low mass-to-light ratios. 
Tidal dwarfs may also have unusual metal abundances
since they were made from material first processed in a much larger galaxy.
Elmegreen \et\ (1993) have also suggested that the age distribution of 
their stellar populations may also be peculiar
since there will be old stars from the spiral along with stars that 
have formed
in a burst of star formation when the dwarf first formed and stars that
have formed at a much slower pace since then.

Some of these properties, like the amount of dark matter, 
will not depend on how long ago the tidal dwarf formed
but others will.
Tidal dwarfs forming out of modern day spirals are
observed to be metal-rich because they have formed out of material already 
processed in a giant spiral
(Duc \& Mirabel 1999, Hunsberger \et\ 2000);
most dwarf irregulars today, on the other hand, are metal
poor compared to all but the extreme outer parts of spiral disks.
However, a tidal dwarf that
formed from a spiral many Gyrs ago 
when spirals were still metal poor
($\geq$6 Gyrs ago for material drawn from the central regions
of a galaxy like M33, $\geq$9 Gyrs ago for a galaxy like
the Milky Way [Moll\'a, Ferrini, \& D\'iaz 1997]).
could have a metallicity today that is consistent with that
of traditional, metal-poor dwarfs.

Another problem comes in
disentangling evolutionary effects from initial conditions.
For example, a galaxy which underwent a strong burst of star formation 
several Gyrs ago might be a tidal dwarf that formed then or it could
be a tiny galaxy with a peculiar star formation history. 
We do know that many irregulars evolve with star formation rates that
vary in amplitude by factors of a few, as would be expected in small galaxies
(Ferraro \et\ 1989; Tosi \et\ 1991; Greggio \et\ 1993;
Marconi \et\ 1995; Gallart \et\ 1996a,b; Aparicio \et\ 1997a,b;
Dohm-Palmer \et\ 1998, Gallagher \et\ 1998, Gallart \et\ 1999).
However, others show evidence of higher amplitude variations either 
currently or in the past, some 
of which may be statistically significant
(Israel 1988, Tolstoy 1996, Dohm-Palmer \et\ 1997,
Greggio \et\ 1998, Tolstoy \et\ 1998).
Why statistically large variations occur in some seemingly isolated irregulars 
and not in others is not clear although
arguments have been made that tiny galaxies should
evolve via periodic starbursts 
(see, for example, Gerola, Seiden, \& Schulman 1980).
Clearly, we do not understand how irregular galaxies evolve well enough
yet to be able to say whether a particular evolutionary history can
only be consistent with a tidal dwarf formation scenario.

With these problems in mind, we examine properties of tidal dwarfs
that are potentially different from those of traditional dwarfs.
We are asking the question: Do any normal Im galaxies that
we know of have properties that are consistent with a tidal dwarf
origin?

Tidal dwarfs observed currently at the ends of tidal tails have
M$_B$ of $-$14 to $-$19 (Mirabel \et\ 1992; Duc \& Mirabel 1994; 
Hibbard \et\ 1994), and a luminosity function of tidal dwarfs in
compact groups extends to as bright as $-$18 in M$_R$ (Hunsberger
\et\ 1996, converted to H$_0=65$ \kms\ Mpc$^{-1}$).
An older tidal dwarf, however, might have faded from the glory days of
formation. Therefore, we expect to find tidal dwarfs with M$_B>-18$.
In Table \ref{tablist} we keep a running list of dwarfs that stand
out in the properties discussed below as possible tidal dwarf candidates.

\subsection{Lack of Dark Matter}

Barnes \& Hernquist (1992) found that two interacting galaxies with
extended massive dark haloes produced tidal dwarfs with $<$5\% of
its mass in dark matter.  (How this property might be effected by
a different distribution of dark matter in the parent galaxies
is not explored.)
This lack of dark matter in tidal dwarfs 
is potentially the most useful
distinguishing feature since that property will not change
with time and it does not depend on when the dwarf formed.
Most disk galaxies have rotation
curves that require the presence of dark matter, and some studies
of irregulars have argued that irregulars are just as dominated by,
or even more dominated by, dark matter than spirals.
Here we look for signs of unusual dark matter properties among irregulars through
rotation curves and the traditional Tully-Fisher relationship.

Rotation curves of irregular galaxies are a mixed bag. There are
those that look like normal disk rotation curves: they rise as a solid
body, peak, and level off or even begin to fall
(for example, DDO 154: Carignan \& Purton 1998).
In other irregulars the rotation curve rises but never peaks,
presumably because the rotation curves have not been
observed far enough out. 

However, there are also irregulars that have been found to have
no measureable rotation with upper limits on V$_{rot,max}$sin $i$
of $\leq$7.5 \kms.
These include DDO 69 ($=$Leo A;  Young \& Lo 1996),
DDO 99, DDO 120, DDO 143 (Swaters 1999),
DDO 155 ($=$GR 8;  Lo, Sargent, \& Young 1993;
but see also Carignan, Beaulieu, \& Freeman [1990] who interpret
the velocity field as having some rotation in the inner 250 pc);
DDO 187 (Swaters 1999), DDO 210,
DDO 216 ($=$Pegasus Dwarf), LGS3 (Lo, Sargent, \& Young 1993);
Sag DIG (Young \& Lo 1997), and NGC 4163 (Swaters 1999).
This is in contrast to low surface brightness spirals that, in
spite of their very low surface brightness levels, nevertheless,
rotate at high speeds compared to irregulars (de Blok, McGaugh,
\& van der Hulst 1996).
Galaxies with no measureable organized rotation may be
good candidates for no dark matter.

The maximum rotation speeds of
irregulars and a sample of spirals taken from Broeils (1992)
are shown in Figure \ref{figvmax}. Upper limits to the rotation speed
are used to put the galaxies with no measureable
rotation on this plot.
One can see that at an M$_B$ of about $-$15 and fainter
the irregular galaxies deviate strongly
from the relationship between M$_B$ and V$_{rot,max}$ determined 
for spirals. The deviation from the relationship is
in the sense that many irregulars have rotation speeds that are too
small for their luminosity. This is in the sense that one expects
for galaxies that have too little dark matter. 

In a study of dark matter in late-type dwarf galaxies,
Swaters (1999) has shown that there may not be a lot of dark matter in
the disks of irregulars, but dark matter in haloes is still required
to explain the rotation curves. 
However, in some galaxies 
the evidence for any dark matter at all is not strong (for example, 
NGC 1569: Stil 1999).
Swaters (1999) found
5 galaxies in his study, including 3 Ims (DDO 50, DDO 125, DDO 143),
for which a maximum disk fit to the rotation curve leaves
no room for dark matter. He
suggests that one of these galaxies, DDO 125, is
a good candidate for a tidal dwarf since it lies near 
giant HI streamers associated with the larger
irregular galaxy NGC 4449 (Hunter \et\ 1998).

In Figure \ref{figtf} we 
consider an alternative means of looking at the mass, and hence
dark matter, in galaxies: the traditional Tully-Fisher relationship
(Tully \& Fisher 1977). 
The Tully-Fisher relationship
is shown in Figure \ref{figtf}a where
we plot log W$_{20}^{ci}$ 
against M$_B^i$. W$_{20}^{ci}$ is the full width at
20\% intensity of the integrated HI profile, corrected for instrumental
broadening, random motions, and the inclination of the galaxy (Broeils 1992).
The inclinations were determined using minor-to-major axis ratios 
from Swaters (1999) and de Vaucouleurs \et\ (1991,$=$RC3) and
assuming an intrinsic ratio of 0.3 (Hodge \& Hitchcock 1966, van den Bergh
1988).  The M$_B^i$ have been corrected for internal absorption,
also dependent on the galaxy inclination, according to Broeils (1992)
with a reduction of a factor of 4 in the absorption compared to
spirals to better match the
observed lower dust contents of irregulars.
W$_{20}^{ci}$ should measure approximately twice the maximum
rotation speed and so should be related to the total mass in the galaxy.
This plot has the advantage over the plot shown in Figure \ref{figvmax}
that integrated HI profiles are available for far more galaxies than
are velocity fields.
We see that
there is much more scatter among the irregular galaxies than the spiral
sample,
but the scatter for the brighter irregulars is distributed
around the relationship defined by the spirals with more falling below
the relationship than above it.
However, 
at the low end of the relationship---low HI widths and low luminosities,
several irregulars are too bright for their HI widths.
This deviation above the relationship 
is consistent with the possibility that those
galaxies have less dark matter than other galaxies of that luminosity
although statistics on low luminosity galaxies are poor.

Carignan \& Beaulieu (1989), Swaters (1999), and McGaugh \et\ (2000)
have observed that 
the Tully-Fisher relationship begins to break down for low luminosity galaxies
although the galaxies in their samples deviate in the sense of being
underluminous for their rotation speed.
Milgrom \& Braun (1988)
suggest that the relationship is maintained if M$_B^i$ is replaced
with the total luminous mass, including gas.
We have examined this possibility in 
Figure \ref{figtf}b.
We have estimated the mass in stars from M$_B^i$ and a
stellar mass-to-light ratio that is appropriate
for a galaxy forming stars at a constant rate for 10 Gyr with a normal
stellar initial mass function (Larson \& Tinsley
1978). This is clearly a rough approximation and will not apply equally
well to all of the galaxies
in the sample.  However, this is most likely to break down
for the dwarf galaxies
for which errors of factors of even 10 in stellar mass will not make very
much difference to the mass in gas plus stars because the masses are 
usually dominated
by the gas.
The mass in gas is HI plus He (1.34$\times$M$_{HI}$) and
does not include molecular gas since this quantity is not known for most
irregulars.
The resulting relationship looks similar to
that in 
Figure \ref{figtf}a, and, if one compares equal logarithmic
intervals, the scatter is no less. 
This is in contrast to what McGaugh \et\ (2000) found:
that galaxies with W$_{20}^{ci}<180$ \kms\ in their sample,
that extends to a mass of $10^7$ M\solar, are brought into
agreement with spirals once the gas content is taken into account.

In Figure \ref{figtf}b we see that only a few
galaxies deviate more than most irregulars. Several, such as
DDO 155, NGC 4163, and the Sm galaxy DDO 135, fall below the
relationship for the rest of the galaxies. This is in the opposite
sense of what we are looking for. Of those galaxies above the
relationship, only IC 1613, DDO 210, DDO 216, and SagDIG, because
they are upper limits (all but IC 1613 have no measureable rotation), have
the potential to fall further outside the relationship than the
bulk of the galaxies.
(We ignore the spiral DDO 80, possibly interacting, discussed by Broeils [1992].)

\subsection{Metallicity}

At the time of formation
the tidal dwarf 
will take on the metallicity of the material drawn from the
parent spiral, and, since spirals
have evolved more rapidly than irregulars,
this could make the tidal dwarf more metal rich (Schweizer 1978). 
In fact, many tidal dwarfs in formation observed today are too metal
rich for their luminosity (Duc \& Mirabel 1997, Hunsberger \et\ 2000).
For normal irregulars, Richer \& McCall (1995) found that the
scatter in the metallicity-luminosity relationship increases for galaxies with
M$_B>-15$. This could imply that
low luminosity galaxies have a more diverse evolutionary background. 
We have examined the position of irregular galaxies on a plot
of oxygen abundance versus
M$_B$, adapted from Hunter \& Hoffman (1999).
In Figure \ref{figoh}
we show the oxygen abundance
plotted against M$_B$ for a sample of spiral and irregular
galaxies along with Richer and McCall's relationship for low-luminosity
galaxies.
The scatter even among the spirals is substantial.
Although
most of the irregulars with no measureable rotation
deviate substantially from the 
relationship, the scatter among all of the galaxies is too high
to be able to say that they deviate more than most.
Clearly the lower luminosity end of this relationship needs to
be explored further.

We have also included a few Im galaxies in the Virgo cluster for comparison.
Potentially the higher density of clusters like Virgo will result in a
higher population of tidal dwarfs as has been found in compact clusters
(Hunsberger \et\ 1996).
V\'ilchez (1995) had suggested that irregulars
in Virgo
are in fact more metal rich than field irregulars, but he
assumed the high metallicity branch of the double-valued
relationship between
emission-line ratio and abundance. 
On the other hand,
Lee, McCall, \& Richer (1998) found
metallicities for Virgo irregulars that are consistent with the relationship.
In our plot, where we assumed the lower branch of the metallicity,
the Virgo Cluster galaxies do not stand out from the general
scatter.

However, there are also complications to using metalliciy as
a tidal dwarf indicator: 1)
If the material that forms the tidal dwarf comes primarily
from the outer part of the spiral, it could be just as metal poor 
as an irregular. 
2) If a tidal dwarf formed many Gyr ago,
the starting metallicity would be lower than if it had formed today.
3) The metallicity will change as the dwarf evolves and how it changes
is convolved with how it evolves.
4) There are substantial observational uncertainties in determining the oxygen
abundance.

\subsection{Structure}

Irregular galaxies are generally believed to be disk systems, although
thicker than spirals
Hodge \& Hitchcock 1966, van den Bergh
1988).  
However, there is some debate about even this
basic property of irregulars (Sung \et\ 1998).
However, we see no reason why a dwarf formed in a tidal tail would
have to be a disk. 
Furthermore,
the irregulars that have no measureable disk rotation could
be ones that are not disk-shaped.
Intriguingly, Patterson \& Thuan (1996) 
examined the surface photometry and scale lengths
of a sample of
dwarf irregular galaxies and found that they divided into two groups.
One group has scale lengths like those of dwarf ellipticals and twice
those of BCDs, and the other is comparable to BCDs and half that of
dEs. 
Could these two groups also be related to the two origins?
At this point, we cannot tell which group would be the tidal
dwarfs.
In addition, studies of the irregulars 
WLM and NGC 3109 have shown
that those irregulars have a halo of old stars in addition to their disk
(Minniti \& Zijlstra 1996; Minniti, Zijlstra, \& Alonso 1999).
They point out that by contrast all of the old stars in the LMC
are in its disk.  However, whether this is normal or abnormal
for irregulars is not yet clear.

\subsection{Stellar Population}

Another possible oddity of a tidal dwarf is its stellar population.
Elmegreen \et\ (1993) argue that the tidal dwarf should consist
of a small fraction ($\leq$40\%) of old stars from the parent spiral,
a strong starburst at formation, and a normal distribution
of mixed ages of stars formed since the galaxy's formation.
Unfortunately, not many color magnitude diagrams of irregulars
go deep enough for analysis to pull out limits on star formation
histories more than a few Gyrs into the past.
Of those that can put some limits up to 10 Gyrs ago,
6 galaxies appear to
have normal star formation histories
(DDO 216: Aparicio, Gallart, \& Bertelli 1997a;
LGS3: Aparicio \et\ 1997b;
IC 1613: Cole \et\ 1999;
NGC 3109: Minniti \et\ 1999;
NGC 6822: Gallart \et\ 1996a;
WLM: Minniti \& Zijlstra 1997).
However,
one galaxy may fit the pattern expected of a tidal dwarf:
DDO 69. Tolstoy \et\ (1998) put a limit of $<$10\% of the total
in a very old stellar population, with the majority of the star
formation taking place within the past 1.5 Gyr.
In the scenario of Elmegreen \et, DDO 69 would have formed
about 2 Gyrs ago.

\section{Discussion}

We have examined properties of a sample of irregular galaxies from
the perspective of features that might distinguish galaxies formed
in tidal interactions at some time shorter than a Hubble time
from those formed from collapse of a primordial gas cloud a Hubble
time ago. We have considered the lack of 
dark matter predicted by models as manifested in rotation speeds
and the Tully-Fisher relationship, the fact that tidal dwarfs may
have formed from enriched material, structure, and peculiar stellar
populations. However, using these features to identify old tidal
dwarfs is currently imprecise.
Abundances and star formation histories are entangled
in other evolutionary and observational effects, and not enough is known
about the amount and location of dark matter and the true
structure of irregulars. 

Nevertheless, we have identified candidates 
for old tidal dwarfs, and they are listed in Table \ref{tablist}. 
We have also listed the distance to the nearest large galaxy.
A little over one-quarter of the galaxies in this list
are in the Local Group.
Eighty-five percent of the galaxies are within 0.5 Mpc of a large galaxy; 
and one
lies near supergiant gas streamers wrapped around a nearby Im galaxy.

Because of the difficulties in identifying
old tidal dwarfs, these galaxies can only be considered candidates 
at this point.
In addition this is not an exhaustive list, and
we have not included representative samples of other groups of dwarfs 
including dwarf ellipticals and
dwarf spheroidals. The peculiar galaxy IZw18, for example, has the
peculiar stellar populations and complex kinematics that might make
it a candidate.

Clearly, it is important to understand the formation and evolutionary
processes of the most common galaxy in the universe: irregular and other
dwarf galaxies.
The fact that irregulars could potentially be formed in more than
one way complicates our ability to interpret the properties of the
galaxies that we see today.
How can we improve our understanding of irregulars so that
differences due to different origins might be more apparent? 
We need to better understand the kinematics and structures of irregular
and dwarf galaxies. This includes the gas and stellar kinematics
and velocity dispersions from which we can infer the distribution 
and amount of dark matter and the stellar structure of the galaxy.
We also need more very deep studies of stellar populations of
irregulars, particularly probing the extremes of galaxy characteristics.
Only once there is a statistically significant sample of star formation
histories can we begin to see trends.
Finally, we need numerical simulations that can show whether interactions
are feasible, perhaps between two unequal mass partners, that can produce 
a tidal dwarf and still leave the parent spiral intact. This is
particularly important for the Local Group system in which we
identify 6 candidate old tidal dwarfs, but the obvious parents are
relatively normal looking spirals.

\acknowledgments

This discussion originated while EWR was a summer student at Lowell
Observatory under the Research Experiences for Undergraduates program
operated by Northern Arizona University and funded by the National
Science Foundation under grant number 9423921.
Support for this work was provided by the Lowell Research Fund
and the Friends of Lowell Observatory.

\clearpage
 
\begin{deluxetable}{lccccc}
\tablecaption{Candidate Old Tidal Dwarfs. \label{tablist}}
\tablewidth{0pt}
\tablehead{
\colhead{} & \colhead{} & \colhead{}
& \colhead{} 
& \multicolumn{2}{c}{Nearest Large Galaxy}  \nl
\colhead{Galaxy}       & \colhead{Rotation\tablenotemark{a}} 
& \colhead{T-F\tablenotemark{b}}  
& \colhead{Stellar Pop}
& \colhead{$\Delta$d (kpc)} & \colhead{$\Delta$V (km/s)} 
} 
\startdata
DDO 43      & N       & Y       & \nodata &  690 &  82 \nl
DDO 50      & Y       & N       & \nodata &  510 & 190 \nl
DDO 52      & N       & Y       & \nodata &  710 &  50 \nl
DDO 69      & Y       & Y       & Y       & \multicolumn{2}{c}{Local Group} \nl
DDO 99      & Y       & Y       & \nodata &  380 &   1 \nl
DDO 120     & Y       & Y       & \nodata &   70 &  73 \nl
DDO 125     & Y       & Y       & \nodata &   41 &  10 \nl
DDO 143     & Y       & N       & \nodata &  320 & 160 \nl
DDO 155     & Y       & N       & N       &   68 &  53 \nl
DDO 165     & N       & Y       & \nodata &  260 &  37 \nl
DDO 187     & Y       & Y       & \nodata & 2500 & 150 \nl
DDO 210     & Y       & Y       & \nodata & \multicolumn{2}{c}{Local Group} \nl
DDO 216     & Y       & Y       & N       & \multicolumn{2}{c}{Local Group} \nl
CVndwA      & Y       & \nodata & \nodata &  250 &  12 \nl
Haro 4      & N       & Y       & \nodata &  170 &  35 \nl
IC 1613     & \nodata & Y       & N       & \multicolumn{2}{c}{Local Group} \nl
LGS3        & Y       & N       & N       & \multicolumn{2}{c}{Local Group} \nl
M81dwA      & \nodata & Y       & \nodata &  500 & 150 \nl
NGC 1569    & Y       & N       & \nodata &  240 & 135 \nl
NGC 4163    & Y       & Y       & \nodata &   95 &  79 \nl
Sag DIG     & Y       & Y       & \nodata & \multicolumn{2}{c}{Local Group} \nl
\enddata
\tablenotetext{a}{Galaxies that deviate significantly from the 
relationship in Figure \ref{figvmax}.}
\tablenotetext{b}{Galaxies that potentially deviate from the 
relationship in Figure \ref{figtf}.}
\end{deluxetable}

\clearpage

\figcaption{Maximum rotation speed plotted against absolute blue magnitude.
Arrows pointing to the right are lower limits to the maximum
rotation speed because
the rotation curve was not observed to level off.
Open squares are upper limits to the rotation speed because no rotation
was measured.
The spiral galaxies are from Broeils (1992), as is the solid line fit to
the spirals (modified for an H$_0$ of 65 \protect\kms\ Mpc$^{-1}$).
The dashed lines are the solid line offset 3.9 magnitudes
in M$_B$, enough to encompass
the bulk of the scatter around the solid line.
M$_B$ for the Im and BCD galaxies was corrected for internal absorption
assuming an E(B$-$V)$_i$ of 0.05. Foreground absorption is taken from
Burstein \& Heiles (1984) with the extinction curve of
Cardelli, Clayton, \& Mathis (1989).
The rotation data for irregulars are taken from 
Sargent, Sancisi, \& Lo (1983);
Comte et al.\ (1986);
Skillman et al.\ (1988);
Carignan \& Beaulieu (1989);
Lake \& Skillman (1989);
Jobin \& Carignan (1990);
Broeils (1992);
Puche et al.\ (1992);
Lo, Sargent, \& Young (1993);
Simpson (1995);
Young \& Lo (1996, 1997);
Broeils \& Rhee (1997);
McIntyre (1997);
van Zee et al.\ (1997);
Wilcots \& Miller (1998);
Stil (1999);
Swaters (1999); and
Hunter, Elmegreen, \& van Woerden (2000).
\label{figvmax}}

\figcaption{W$_{20}^{ci}$ is the full width of the integrated
HI profile at 20\% intensity, corrected for instrumental broadening,
random motions, and inclination of the galaxy
(Bottinelli et al.\ 1990, Broeils 1992).
W$_{20}^{ci}$ is approximately twice the maximum rotation speed
of the HI gas in the galaxy and so is related to the total mass
of the galaxy.
Those galaxies for which the correction for random motions is larger
than the observed W$_{20}$ are assigned a W$_{20}^{c}$ of 5 km s$^{-1}$
and then corrected for inclination. They are
shown as upper limits.
a) Upper panel: 
The solid line is the relationship for spirals from Broeils (1992)
converted to an H$_0$ of 65 km s$^{-1}$ Mpc$^{-1}$.
b) Bottom panel:
The mass in gas is taken to be 1.34 times the mass in HI, to account for He.
The mass in stars is 1.54 times the blue luminosity in solar units.
The solid line is our least squares fit to the spirals.
The dashed lines are the solid line offset 0.7 in the logarithm
of the mass, enough to encompass
the bulk of the scatter around the solid line.
\label{figtf}}

\figcaption{Figure adapted from Hoffman \& Hunter (1999) with a few
Virgo irregulars
and irregulars with no measureable rotation included.
Data for the Virgo irregulars are taken from Gallagher \& Hunter
(1989) and V\'ilchez (1995) and rederived
with the assumption that the oxygen abundance is the lower of
the two possibilities for the emission-line ratios.
The solid line is the relationship
derived by Richer \& McCall (1995)
for low luminosity galaxies.
The dashed lines are the solid line offset 0.4 dex in log O/H$+$12,
enough to encompass
the bulk of the scatter around the solid line.
Spirals are from
Zaritsky, Kennicutt, \& Huchra (1994).
Except for the spirals,
galaxies are distinguished in the plots by the method used to
estimate O/H:
The label ``4363'' means that the determination of O/H was from use
of [OIII]$\lambda$4363 to determine T$_e$,
``McG'' refers to the method of McGaugh(1991),
and ``lit'' refers to values taken from the literature.
We have left off measurements determined by 
the method of Edmunds \& Pagel (1984) that has a higher uncertainty.
M$_B$ for the non-spiral galaxies was corrected for internal absorption
assuming an E(B$-$V)$_i$ of 0.05. Foreground absorption is taken from
Burstein \& Heiles (1984).
\label{figoh}}

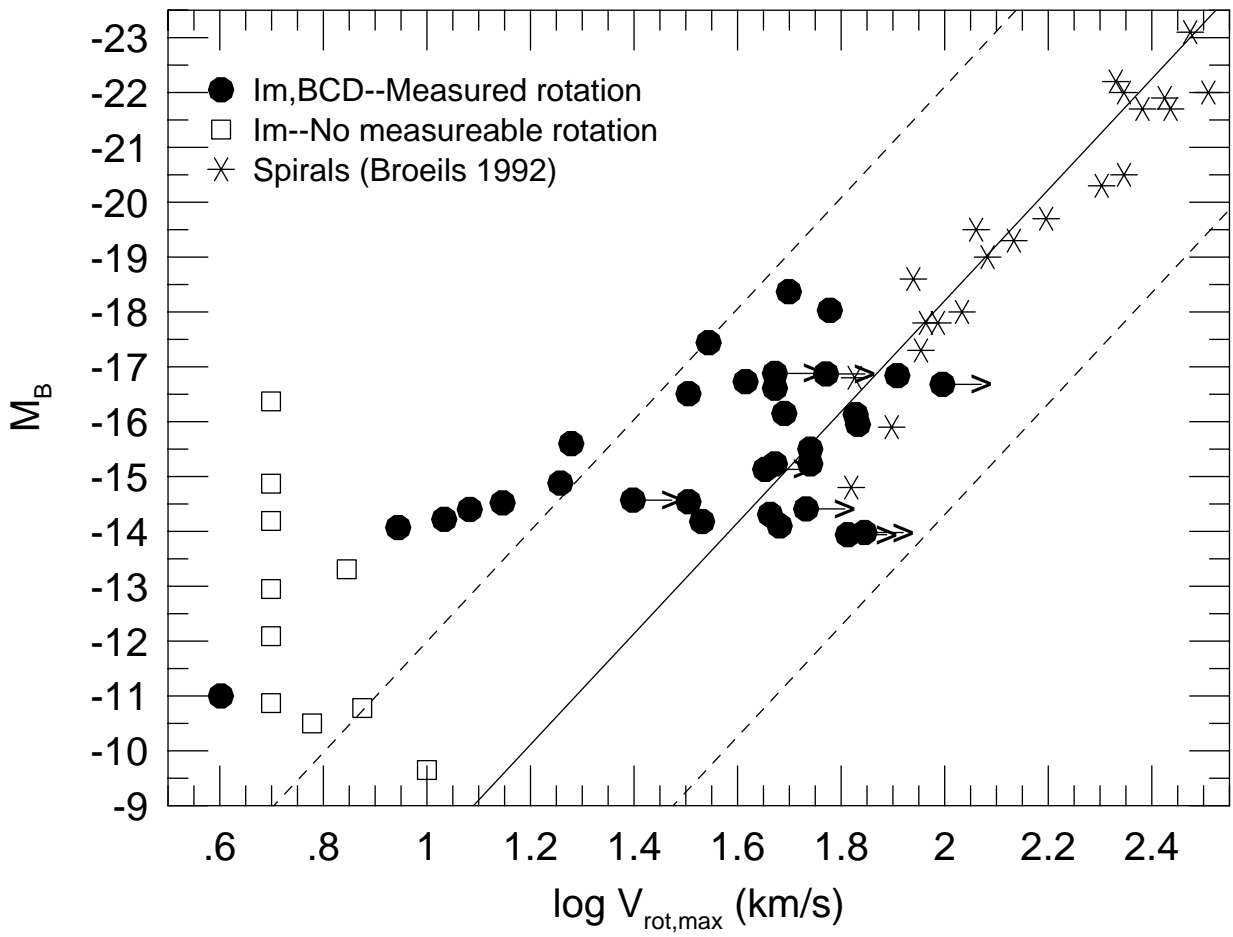
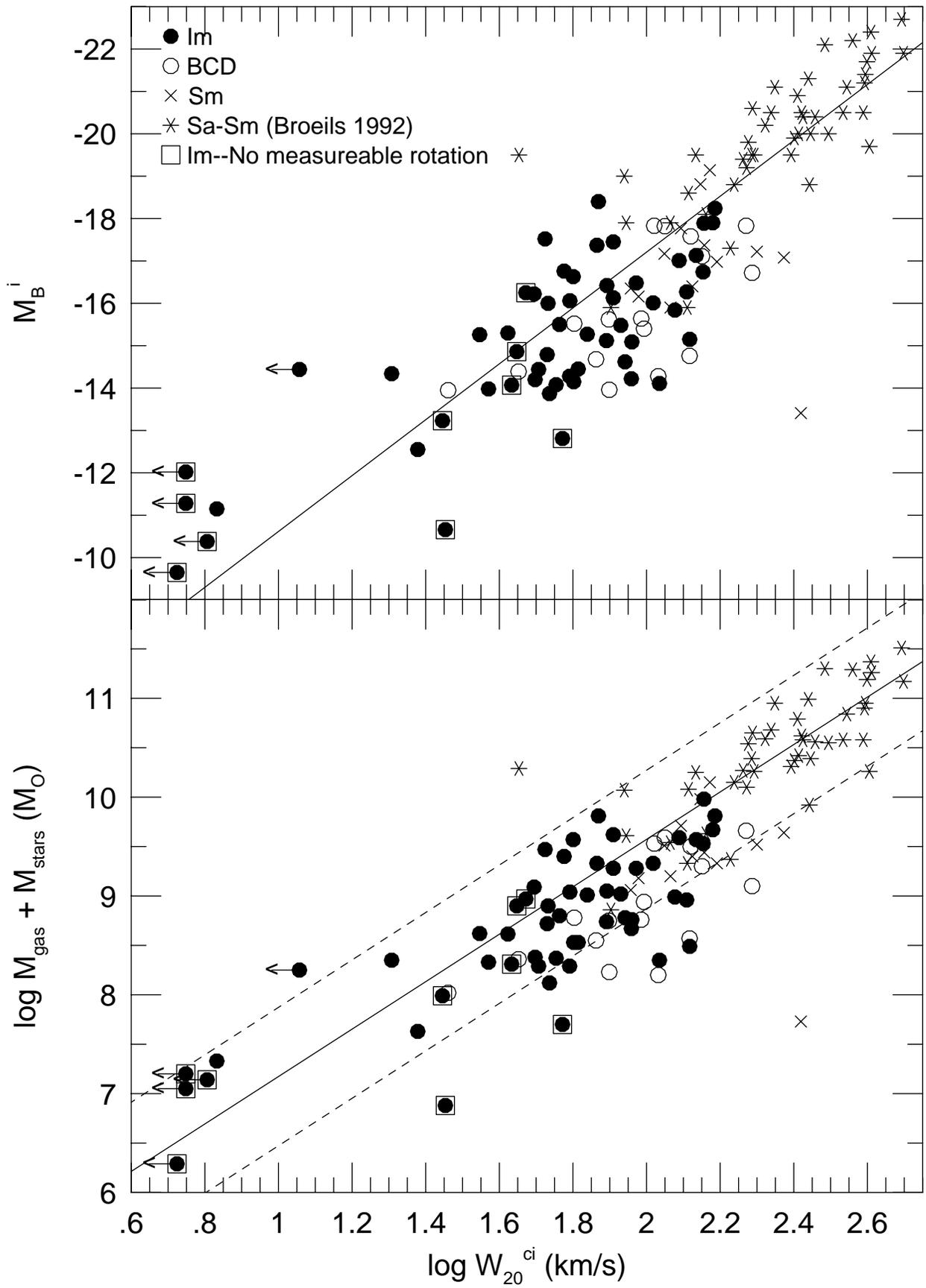
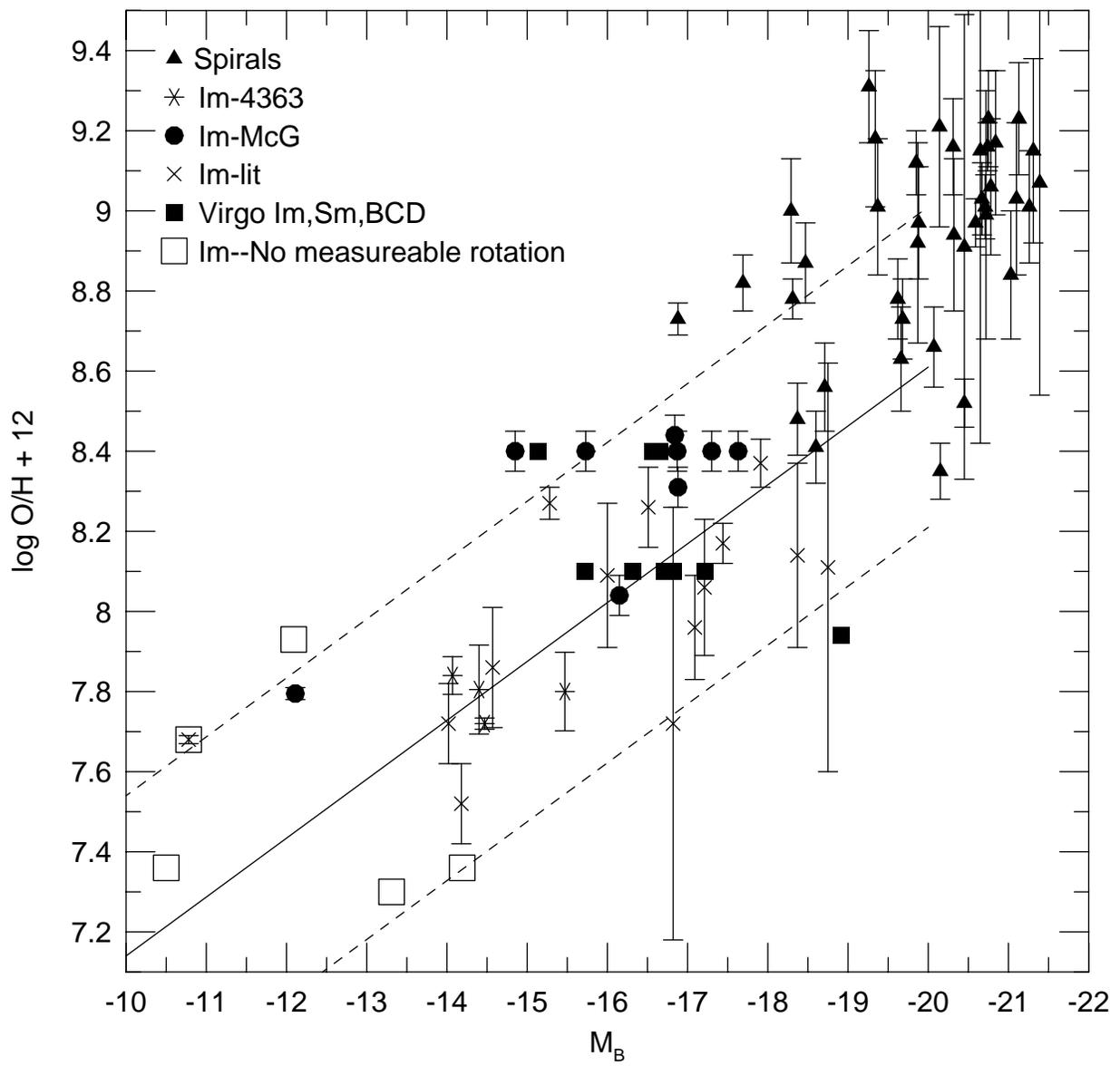
\end{document}